\documentclass[12pt]{article}
\usepackage{graphicx}
\usepackage{bm}
\usepackage[utf8]{inputenc}
\usepackage{lmodern}
\begin{document}
\title{Nonlinear low-frequency collisional quantum Buneman instability} 
\date{}
\maketitle 
\begin{center}
\large{F. Haas 
\vskip.3cm
Departamento de F{\'i}sica, Universidade Federal do Paran\'a, 81531-990, Curitiba, Paran\'a, Brazil

\vskip.5cm
A. Bret
\vskip.3cm
ETSI Industriales, Universidad de Castilla-La Mancha, 13071 Ciudad Real, Spain and
   Instituto de Investigaciones Energ\'eticas y Aplicaciones Industriales, Campus Universitario de Ciudad Real, 13071 Ciudad Real, Spain}
   \end{center}
   
   \vskip.3cm
\begin{abstract}{The Buneman instability occurring when an electron population is drifting with respect to the ions is analyzed in the quantum linear and nonlinear regimes. The one-dimensional low-frequency and collisional model of Shokri and Niknam [Phys. Plasmas, \textbf{12} (2005) 062110] is revisited introducing the Bohm potential term in the momentum equation. The linear regime is investigated analytically, and quantum effects result in a reduction of the instability. The nonlinear regime is then assessed both numerically and analytically, and pure quantum density oscillations are found to appear during the late evolution of the instability.}
\end{abstract}
\vskip.3cm

The Buneman instability \cite{Buneman1959} is a basic ins\-ta\-bi\-li\-ty process in classical plasmas. It occurs in beam/plasma systems when there is a significant drift between electrons and ions. It is frequently referred to as the ``Farley-Buneman'' instability because it was almost simultaneously discovered by Farley \cite{Farley1963}. In view of the very basic ``set-up'' it needs to be triggered, this instability has been found to play a role in many physical scenarios. In space physics and geophysics, this instability has been invoked in the earth ionosphere \cite{Farley1963} or in the solar chromosphere \cite{Gogoberidze}. In double-layer and collisionless shock physics, the same instability has been found responsible, under certain circumstances, of the very formation of this kind of structures \cite{Iisuka1979}. In si\-tua\-tions where an electron beam enters a plasma, like in the fast ignition scenario for inertial fusion \cite{Tabak1994}, the electronic return current prompted in has been found to generate exponentially growing Buneman modes through its interaction with the plasma ions \cite{Lovelace, Bret2008}.

On the other hand, presently quantum plasmas are attracting much attention from a variety of reasons, since they provide a typical state of ionized matter under large densities and/or low temperatures. For instance, quantum effects in plasmas are relevant in ultra-small semiconductor devices, metal clusters, intense laser-solid interaction experiments and compact astrophysical objects like neutron stars and white dwarfs (see \cite{rmp, usp} for reviews). Moreover, X-ray Thomson scattering techniques in dense plasmas have been used \cite{glenzer} to verify the signature of quantum diffraction effects in the dispersion relation for e\-lec\-tros\-ta\-tic waves. Also, in the near future the development of coherent brilliant X-ray radiation sources \cite{thiele} and keV free electron lasers \cite{gregori} will provide experimental access to the quantum nature of plasmas under extreme conditions. Some of the most recent developments in the field are the analysis of wave breaking in quantum plasmas \cite{bleker}, the characteristics
of bounded quantum plasmas including electron exchange-correlation
effects \cite{Ma},  the development of a quantum single-wave theory for nonlinear coherent structures in quantum plasmas \cite{tzenov}, the discussion of waves in quantum dusty plasmas \cite{Stenflo}, the prediction of a fundamental size limit for plasmonic devices due to the quantum broadening of the transition layer \cite{marklund},
as well as the inclusion of spin \cite{brodin} and relativistic \cite{tito} effects in quantum plasma modeling. Finally, we note the usefulness of quantum plasma techniques to other, closely related problems, such as the treatment of nonlinear wave propagation in gravitating Bose-Einstein condensates \cite{ghosh}.

The aim of the present work is to discuss the quantum analog of the Buneman instability. Our approach is based on the quantum hydrodynamic model for plasmas, which has proven to be very useful for the understanding of nonlinear problems in quantum Coulomb systems \cite{book}. More specifically, we consider linear and nonlinear low-frequency waves in a collisional electron-ion plasma, extending the model by Shokri and Niknam \cite{shokri} by means of the inclusion of a quantum term (the so-called Bohm potential) associated to the wave nature of the quantum particles. As verified in the continuation, the Bohm potential has a stabilizing influence on the low-frequency linear Buneman instability. In addition, it eventually produces nonlinear oscillatory structures at the late stages of the instability, in sharp contrast to the monotonic character of the classical stationary states.

We start with the two-species cold quantum hydrodynamic model \cite{ms}, taking into account the effect of collisions using simple relaxation terms,
\begin{eqnarray}
\label{ee1}
\frac{\partial n_e}{\partial t} + \frac{\partial (n_e v_e)}{\partial x} &=& 0 \,,\\
\frac{\partial n_i}{\partial t} + \frac{\partial (n_i v_i)}{\partial x} &=& 0 \,,\\
\frac{\partial v_e}{\partial t} + v_e \frac{\partial v_e}{\partial x} &=& - \frac{e E}{m}- \nu_e v_e\\
 && + \frac{\hbar^2}{2m^2}\frac{\partial}{\partial x}\left(\frac{\partial^{2}\sqrt{n_e}/\partial x^2}{\sqrt{n_e}}\right)\,,\nonumber\\
\frac{\partial v_i}{\partial t} + v_i \frac{\partial v_i}{\partial x} &=&  \frac{e E}{M} - \nu_i v_i \,,\\
\label{ee5}
\frac{\partial E}{\partial x} &=& \frac{e}{\varepsilon_0} (n_i - n_e) \,.
\end{eqnarray}
In Eqs. (\ref{ee1})--(\ref{ee5}), $n_{e,i}$ are the electron (ion) number densities, $v_{e,i}$ the electron (ion) fluid velocities, $E$ the electrostatic field, m (M) the electron (ion) mass, $-e$ the electron charge, $\hbar$ the Planck constant divided by $2\pi$ and $\varepsilon_0$ the vacuum permittivity. Moreover, $\nu_{e,i}$ represent electron (ion) collision frequencies with neutrals. For simplicity in this work only one spatial dimension is considered. Quantum effects are included in the force equation for electrons by means of the $\sim \hbar^2$ term, the so-called Bohm potential. Due to $m/M \ll 1$, no quantum terms are needed in the ion force equation.

Linearizing the model around the homogeneous e\-qui\-li\-brium
\begin{equation}
n_{e,i} = n_0 \,, \quad v_e = -\frac{e E_0}{m\nu_e} \,, \quad v_i = \frac{e E_0}{M\nu_i} \,, \quad E = E_0 \,,
\end{equation}
where $E_0$ is an external DC electric field. Supposing perturbations $\sim \exp(i[kx-\omega t])$ with wavenumber $k$ and wave frequency $\omega$, the result in the reference frame of the drifting ions is
\begin{eqnarray}
1 &-& \frac{\omega_{pe}^2}{(\omega - kv_0) (\omega - kv_0 + i\nu_e) - \hbar^2 k^4/(4m^2)}\nonumber\\
&-& \frac{\omega_{pi}^2}{\omega (\omega + i\nu_i)} = 0 \,,
\end{eqnarray}
where $\omega_{pe} = (n_0 e^2/(m\varepsilon_0))^{1/2}$ and $\omega_{pi} = (n_0 e^2/(M\varepsilon_0))^{1/2}$ are resp. the electron and ion plasma frequencies and where
\begin{equation}
v_0 = - e E_0 \left(\frac{1}{m\nu_e} + \frac{1}{M\nu_i}\right)
\end{equation}
is the relative electron-ion equilibrium drift velocity.

In the very low frequency range $\omega \ll \nu_i \ll kv_0$, $\nu_e \ll kv_0$, the dispersion relation reduces to
%
%
%
\begin{equation}
\label{dr}
\omega = \frac{i\omega_{pi}^2}{\nu_i} \frac{k^2 (v_{0}^2- \hbar^2 k^2/(4 m^2))}{\omega_{pe}^2 - k^2 v_{0}^2 + \hbar^{2}k^4/(4m^2)} \,.
\end{equation}
This mode is unstable (${\rm Im}(\omega) > 0$) provided
\begin{equation}
\label{sp}
\omega_{pe}^2 > k^2 v_{0}^2 - \frac{\hbar^2 k^4}{4m^2} > 0 \,,
\end{equation}
otherwise it is damped. Small wavelengths such that $\hbar^2 k^2 > 4m^2 v_{0}^2$ are automatically stable, due to the quantum effects.

Assuming a large ion-neutral collision frequency where $\nu_i \gg \omega_{pi}$, we can suppose a slow temporal dynamics. Moreover, from Eq. (\ref{sp}) we have  $v_0/\omega_{pe}$ as a natural choice of spatial scale for the development of the instability, at least for not very large quantum effects. Therefore, we consider the following rescaling,
\begin{eqnarray}
t &\rightarrow& \frac{\omega_{pi}^2 t}{\nu_i} \,, \quad x \rightarrow \frac{\omega_{pe} x}{v_0} \,, \quad
v_e \rightarrow \frac{v_e}{v_0} \,, \\ v_i &\rightarrow& \frac{M \nu_i v_i}{m \omega_{pe} v_0} \,, \quad 
n_{e,i} \rightarrow \frac{n_{e,i}}{n_0} \,, \quad E \rightarrow \frac{e E}{m \omega_{pe} v_0} \,. \nonumber
\end{eqnarray}

For simplicity using the same symbols for original and transformed variables, the rescaled system reads
\begin{eqnarray}
\frac{\omega_{pi}^2}{\nu_i \omega_{pe}} \frac{\partial n_e}{\partial t} + \frac{\partial (n_e v_e)}{\partial x} &=& 0 \,,\\
\frac{\partial n_i}{\partial t} + \frac{\partial (n_i v_i)}{\partial x} &=& 0 \,,\\
\frac{\omega_{pi}^2}{\nu_i \omega_{pe}} \frac{\partial v_e}{\partial t} + v_e \frac{\partial v_e}{\partial x} &=& - E - \frac{\nu_e}{\omega_{pe}} v_e \\
&& + \frac{H^2}{2}\frac{\partial}{\partial x}\left(\frac{\partial^{2}\sqrt{n_e}/\partial x^2}{\sqrt{n_e}}\right) \,,\nonumber\\
\frac{\omega_{pi}^2}{\nu_{i}^2} \left(\frac{\partial v_i}{\partial t} + v_i \frac{\partial v_i}{\partial x}\right) &=&  E - v_i \,,\\
\frac{\partial E}{\partial x} &=& n_i - n_e \,,
\end{eqnarray}
where
\begin{equation}
H = \frac{\hbar\omega_{pe}}{mv_{0}^2}
\end{equation}
is a non-dimensional parameter measuring the relevance of the Bohm potential.

Provided
\begin{equation}
\nu_i \gg \omega_{pi} \,, \quad \omega_{pe} \gg \nu_e \,,
\end{equation}
we obtain
\begin{eqnarray}
\label{e1}
\frac{\partial (n_e v_e)}{\partial x} &=& 0 \,,\\
\frac{\partial n_i}{\partial t} + \frac{\partial (n_i v_i)}{\partial x} &=& 0 \,,\\
\label{bohm}
v_e \frac{\partial v_e}{\partial x} &=& - E\!+\!\frac{H^2}{2}\frac{\partial}{\partial x}\left(\frac{\partial^{2}\sqrt{n_e}/\partial x^2}{\sqrt{n_e}}\right)\!,\\
E &=&  v_i \,,\\ \label{e5}
\frac{\partial E}{\partial x} &=& n_i - n_e \,,
\end{eqnarray}
where the remaining terms are assumed to be of the same order. The final equations are the same as Eqs. (2)--(6) of ref. \cite{shokri}, with the inclusion of the extra $\sim H^2$ contribution. The purpose of the present work is to investigate the role of this quantum term.

We linearize Eqs. (\ref{e1})--(\ref{e5}) around $n_e = n_i = 1, v_e = 1, v_i = 0, E = 0$. Note that after rescaling the equilibrium ion velocity and electric field are higher-order terms, due to $\nu_{e}/\omega_{pe} \ll 1$. We get the dispersion relation
\begin{equation}
\label{wg}
\omega = i\gamma \,, \quad \gamma = \frac{k^2 (1 - H^2 k^2/4)}{1 - k^2 + H^2 k^4/4} \,,
\end{equation}
which is the same as Eq. (\ref{dr}), in terms of non-dimensional variables. Moreover in Eq. (\ref{wg}) we define $\gamma$, the imaginary part of the frequency.

\begin{figure}
\includegraphics[width=0.45\textwidth]{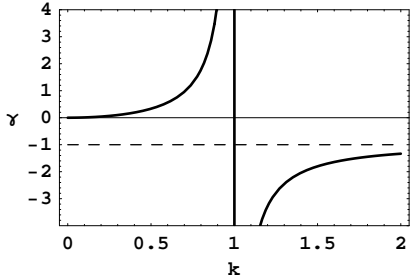}
\caption{Growth rate $\gamma$ in Eq. (\ref{wg}) as a function of the wavenumber $k$, in the classical limit ($H = 0$). Dimensionless variables are used. Note the asymptote at $k = 1$. In addition, $\gamma \rightarrow -1$ as $k \rightarrow \infty$. Instability is found for $0 < k < 1$.}
\label{fig1}
\end{figure}

It is interesting to analyze the behavior of $\gamma$ according to the quantum parameter $H$. In the classical $H = 0$ case, one has $\gamma = k^2/(1-k^2)$ and linear instability for $k < 1$. For the sake of comparison with the non-vanishing $H$ case, in Fig. \ref{fig1} we show the corresponding form of the classical linear instability. The asymptote at $k = 1$ points for an explosive instability. However, this singularity is eventually regularized by nonlinear effects, as discussed in \cite{shokri}.

\begin{figure}
\includegraphics[width=0.45\textwidth]{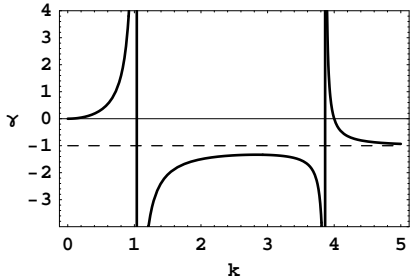}
\caption{Growth rate $\gamma$ in Eq. (\ref{wg}) as a function of the wavenumber $k$, in the semi-classical case ($0 < H < 1$). The value $H = 0.5$ and dimensionless variables were used. Note the asymptotes at $k = k_{A,B}$. In addition, $\gamma \rightarrow -1$ as $k \rightarrow \infty$. Instability is found for $0 < k < k_A$ and also in the small wavelength region $k_B < k < k_C$.}
\label{fig2}
\end{figure}

In the semiclassical $ 0 < H < 1$ case, the growth rate from Eq. (\ref{wg}) has two asymptotes at $k_{A,B}$ defined by
\begin{eqnarray}
\label{kab}
k_{A}^2 &=& \frac{2}{H^2}\left(1 - [1-H^2]^{1/2}\right),\nonumber\\
k_{B}^2 &=& \frac{2}{H^2}\left(1 + [1-H^2]^{1/2}\right).
\end{eqnarray}
Moreover, one has $\gamma > 0$ for $0 < k < k_A$ or $k_B < k < k_C$, where
\begin{equation}
\label{kc}
k_{C}^2 = k_{A}^2 + k_{B}^2 = \frac{4}{H^2} \,.
\end{equation}
By coincidence, one has formally the same instability condition as for the quantum two-stream instability described by a quantum Dawson model, see Eqs. (37)--(42) of ref. \cite{ms}. We note the instability of small wavelengths where $k_B < k < k_C$ has no classical counterpart. The behavior of $\gamma$ as a function of the wavenumber in the semiclassical situation when $0 < H < 1$ is shown in Fig. \ref{fig2}. Again, $\gamma \rightarrow -1$ as $k \rightarrow \infty$.

\begin{figure}
\includegraphics[width=0.45\textwidth]{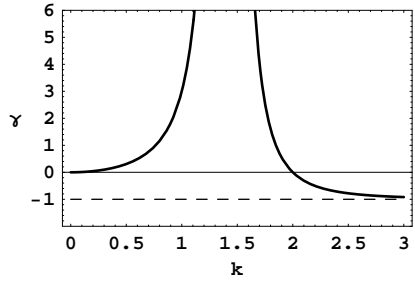}
\caption{Growth rate $\gamma$ in Eq. (\ref{wg}) as a function of the wavenumber $k$, in the particular case $H = 1$. Dimensionless variables are used. Note the asymptote at $k = k_A = k_B = \sqrt{2}$. In addition, $\gamma \rightarrow -1$ as $k \rightarrow \infty$. Instability is found for $0 < k < 2$.}
\label{fig3}
\end{figure}

The case $H = 1$ is particular because then $k_A = k_B = \sqrt{2}$, so that the mid stable branch in Fig. \ref{fig2} disappears. One still has explosive instability, at $k = \sqrt{2}$. The perturbation is linearly stable for $k \geq k_C = 2$. This is shown in Fig. \ref{fig3}.

\begin{figure}
\includegraphics[width=0.45\textwidth]{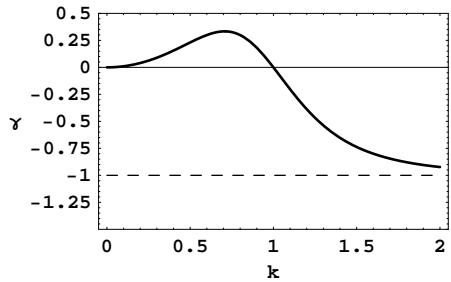}
\caption{Growth rate $\gamma$ in Eq. (\ref{wg}) as a function of the wavenumber $k$, in the strongly quantum ($H > 1$) case. The value $H = 2$ and dimensionless variables were used. Again, $\gamma \rightarrow -1$ as $k \rightarrow \infty$. Instability is found for $0 < k < k_C = 2/H$.}
\label{fig4}
\end{figure}

When $H > 1$, the denominator in Eq. (\ref{wg}) can be shown to be always positive, so that singularities are ruled out. Instability is found for $k < k_C = 2/H$, with the most unstable wavenumber being $k = \sqrt{2}/H$. The corresponding maximal growth rate is $\gamma_{\rm max} = 1/(H^2-1)$. We see that both the unstable $k-$region and the maximal growth rate shrinks to zero as $H$ increases, which is a signature of the ultimate stabilizing nature of the quantum effects. The corresponding function $\gamma(k)$ is shown in Fig. \ref{fig4}.

The overall situation can be visualized in Fig. 5, where the unstable region in $(k^2,H^2)$ space is shown. This is formally the same as Fig. 1 of ref. \cite{ms} on the quantum two-stream instability. On the other hand, for the low-frequency collision-dominated Buneman instability one has distinct behaviors of $\gamma(k)$, according to the parameter $H$. Namely, one has the four classes shown in Figs. \ref{fig1} to \ref{fig4}. In contrast, the quantum two-stream instability exhibits no singularities of the growth rate at specific wavenumbers.

\begin{figure}
\includegraphics[width=0.45\textwidth]{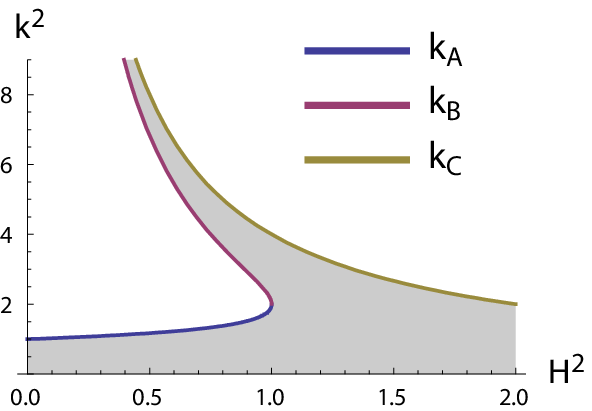}
\caption{Stability diagram for the low-frequency collision-dominated quantum Buneman instability. The filled area is unstable.
Lower and middle curves: resp. $k_{A}^2$ and $k_{B}^2$ as defined in Eq. (\ref{kab}). Upper curve: $k_{C}^2$ as defined in Eq. (\ref{kc}).}
\label{fig5}
\end{figure}

The linear instabilities just described are eventually killed by nonlinear effects. In this context, Eqs. (\ref{e1})--(\ref{e5}) provide a convenient framework for nonlinear studies of the collision-dominated low-frequency quantum Buneman instability. We define the new variables
\begin{equation}
\label{v}
V = \frac{1}{n_e} \,, \quad \rho = n_i - n_e \,,
\end{equation}
giving resp. the electron fluid velocity and net charge density. Equations (\ref{e1})--(\ref{e5}) can then be shown to reduce to
\begin{eqnarray}
\label{ee6}
\rho &=& - \frac{1}{2}\frac{\partial^2}{\partial x^2}\left[V^2 - H^2 \frac{\partial^{2}\sqrt{1/V}/\partial x^2}{\sqrt{1/V}}\right] \,,\\
\label{e6}
0 &=&\frac{\partial V}{\partial t} + V^2 \frac{\partial^2 V}{\partial x^2} \\
&-& \frac{H^2 V^2}{2}\frac{\partial}{\partial x}
\left(\frac{1}{V}\frac{\partial}{\partial x}\left[\frac{\partial^{2}\sqrt{1/V}/\partial x^2}{\sqrt{1/V}}\right]\right) \nonumber\\
&-&\!V^2\!\left(\frac{\partial\rho}{\partial t}\!-\! \frac{\partial}{\partial x}\left[\frac{\rho}{2}\frac{\partial}{\partial x}\left(V^2 - H^2 \frac{\partial^{2}\sqrt{1/V}/\partial x^2}{\sqrt{1/V}}\right)\right]\right)\!.\nonumber
\end{eqnarray}

Assuming quasineutrality ($\rho = 0$) and $V = 1$ at early times, we find the first three terms in Eq. (\ref{e6}) to be initially
the more relevant. Linearizing them we get
\begin{equation}
\label{dif}
\frac{\partial V}{\partial t} = - \frac{\partial^2 V}{\partial x^2} - \frac{H^2}{4}\frac{\partial^4 V}{\partial x^4} \,,
\end{equation}
a quantum-modified diffusion equation with a negative diffusion coefficient. Growing in time solutions are easily found, with the corresponding instability described by the dispersion relation
\begin{equation}
\omega = i k^2 \left(1 - \frac{H^2 k^2}{4}\right) \,.
\end{equation}
This is the quasineutral version of Eq. (\ref{wg}). However, the instability in time is accompanied by a periodic structure in space, a feature not present in the classical case. Indeed, oscillatory in $x$ solutions to Eq. (\ref{dif}) can be also readily constructed {\it e.g.} by separation of variables. In addition, again quantum effects provide the stabilization of the small wavelengths such that $k > k_C = 2/H$.

After the initial increase of the perturbation and enlargement of the density gradient, the $\sim \rho$ terms in Eq. (\ref{e6})
become essential. To examine the stationary states of the model, we set all time-derivatives to be zero. A little algebra then shows that
\begin{equation}
\label{e7}
\frac{d^2}{dx^2}\left(\frac{V^2}{2} - \frac{H^2}{2} \frac{d^{2}\sqrt{1/V}/dx^2}{\sqrt{1/V}}\right) = \frac{1}{V} \,,
\end{equation}
assuming symmetric solutions so that $V'(0)=0$ and excluding the case of identically va\-ni\-shing electric fields.

We start solving Eq. (\ref{e7}) in the classical ($H = 0$) limit. Defining
\begin{equation}
K = \frac{V^2}{2}
\end{equation}
one obtain the Newton-like equation
\begin{equation}
\label{e8}
\frac{d^2 K}{dx^2} = \frac{1}{\sqrt{2K}} \,.
\end{equation}
Assuming $n_{e}(0) = 1$ and a symmetric density profile so that $n_{e}'(0) = 0$, one can integrate Eq. (\ref{e8}) twice with $K(0) = 1/2, K'(0) = 0$. In terms of the electron fluid velocity $V$ the result is
\begin{equation}
(V -1)^{1/2} (V + 2) = X \,, \quad X \equiv \frac{3\, x}{\sqrt{2}} \,,
\end{equation}
which is equivalent to a cubic equation for $V$ with only one real root, namely
\begin{equation}
\label{e9}
V = - 1 + Q + 1/Q \,,
\end{equation}
where
\begin{equation}
Q = \left(\frac{2 + X^2 + \sqrt{X^4 + 4X^2}}{2}\right)^{1/3} \,.
\end{equation}
This is the classical nonlinear stationary Buneman solution, in full agreement with Ref. \cite{shokri}. We note that the ion profile follows from $v_i = E$, with $E$ given by Eq. (\ref{bohm}), as well as from $n_i = \rho + 1/V$, with $\rho$ given by Eq. (\ref{ee6}). It turns out that $n_i \equiv 0$, which is a result from the low-frequency assumption. Indeed, the continuity equation for ions imply $n_i E = cte.$, set to zero in view of the larger ion mass, which in turn imply a negligible stationary ion density. On the other hand, using Eq. (\ref{e9}) one obtain an electric field $E \sim - (6x)^{1/3}$ for large $|x|$, arising from the electron fluid bunching. Also observe that more general solutions are possible to find, including ion density corrections as well as traveling wave forms. However, in this case the algebra becomes much more involved.

Incidentally, we note that from Eq. (\ref{e9}) we have
\begin{equation}
\label{ic}
n_{e}(0) = 1 \,, \quad n_{e}'(0) = 0 \,, \quad n_{e}''(0) = - 1 \,, \quad n_{e}'''(0) = 0 \,,
\end{equation}
a result to be used in the $H \neq 0$ case.

In the quantum case, we need to return to Eq. (\ref{e7}). Defining
\begin{equation}
A \equiv \sqrt{n_e} = \sqrt{1/V} \,,
\end{equation}
one obtain the fourth-order ordinary differential equation
\begin{equation}
\label{e10}
H^2 \frac{d^2}{dx^2}\left(\frac{d^{2}A/dx^2}{A}\right) - \frac{d^2}{dx^2}\left(\frac{1}{A^4}\right) + 2A^2 = 0 \,,
\end{equation}
describing the final, stationary nonlinear stage of the density modulations. For the sake of comparison, we assume
\begin{equation}
\label{e11}
A(0) = 1 \,, A'(0) = 0 \,, A''(0) = -1/2 \,,  A'''(0) = 0 \,,
\end{equation}
which correspond to the same boundary conditions (\ref{ic}) of the classical solution. In other other words, the classical solution is used to set also the second and third-order derivatives of the density at $x=0$, which are needed in the quantum case.

\begin{figure}
\includegraphics[width=0.45\textwidth]{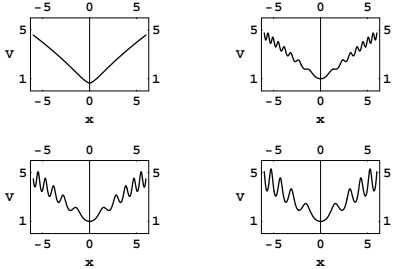}
\caption{Stationary electron fluid velocity. Upper, left: classical exact solution from Eq. (\ref{e9}). The remaining comes from the numerical solution of Eq. (\ref{e10}). Upper, right: $H^2 = 0.5$. Bottom, left: $H^2 = 1.0$. Bottom, right: $H^2 = 1.5$.}
\label{fig6}
\end{figure}

\begin{figure}
\includegraphics[width=0.45\textwidth]{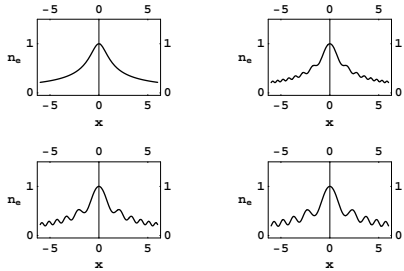}
\caption{Stationary electron fluid density. Upper, left: classical exact solution from Eq. (\ref{e9}). The remaining comes from the numerical solution of Eq. (\ref{e10}). Upper, right: $H^2 = 0.5$. Bottom, left: $H^2 = 1.0$. Bottom, right: $H^2 = 1.5$.}
\label{fig7}
\end{figure}

Equation (\ref{e10}) can be numerically solved, yielding periodic modulations of the stationary velocity and density profiles. The amplitude of the modulations is seen to increase with the strength of the quantum effects, as apparent from Figs. \ref{fig6} and \ref{fig7}. The emergence of new oscillatory structures is ubiquitous in quantum plasmas. In the present case, the ultimate role of the Bohm potential term in Eq. (\ref{bohm}) is a qualitative modification of basic  e\-qui\-li\-brium macroscopic properties like the electron fluid density and velocity. Similar oscillatory patterns of a pure quantum origin appear in the description of weak shocks in quantum plasmas \cite{bychkov}, of the quantum Harris sheet solution in magnetized quantum plasmas \cite{epl} and in undulations of the equilibrium Wigner function in quantum plasma weak turbulence \cite{pre}. 

Considering the extreme quantum case where only the $\sim H^2$ term is taken into account in Eq. (41) can be ins\-truc\-ti\-ve for the physical interpretation of the influence of the Bohm potential. In this $H^2 \gg 1$ situation, one has $A =\cos(x/\sqrt{2})$ in view of the initial conditions in Eq. (42). This imply a density $n_e = A^2$ oscillating with a wavelength $\lambda = \sqrt{2}\,\pi$, of the order of $v_0/\omega_{pe}$ using dimensional coordinates. The behavior so described is confirmed by numerical simulations, with corrections arising from the classical terms. For intermediate values of $H$, numerical analysis shows the wavelength displayed in Figs. \ref{fig6} and \ref{fig7} is actually not constant, but falls like $\sim x^{-3/4}$ instead.

\begin{figure}
\includegraphics[width=0.45\textwidth]{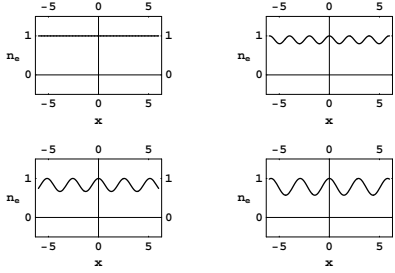}
\caption{Stationary electron fluid density in the quasineutral case, as described by Eq. (\ref{qn}). Upper, left: $H^2 = 0$. Upper, right: $H^2 = 0.5$. Bottom, left: $H^2 = 1.0$. Bottom, right: $H^2 = 1.5$.}
\label{fig8}
\end{figure}

Turning attention to the quasineutral regime, we examine the nonlinear stationary states when $\rho \equiv 0$. In this case, Eqs. (\ref{ee6})--(\ref{e6}) reduce to Eq. (\ref{ee6}) only, the other one being redundant. After twice integrating using Eqs. (\ref{e11}), the model can be shown to be equivalent to the autonomous Pinney's \cite{Pinney} equation
\begin{equation}
\label{pi}
H^2 \frac{d^2 A}{dx^2} + \left(1 + \frac{H^2}{2}\right)A = \frac{1}{A^3} \,.
\end{equation}
Pinney's equation is endemic in nonlinear analysis and is well-known to be exactly solvable. This is specially true in the present autonomous case, where Eq. (\ref{pi}) can be directly integrated twice. Assuming $A(0) = 1, A'(0) = 0$ as before, the solution reads
\begin{equation}
\label{qn}
A^2 = n_e = \frac{1}{1+H^2/2}\left(1 + \frac{H^2}{2}\cos^2\left[\sqrt{1+H^2/2}\,\,\frac{x}{H}\right]\right) \,,
\end{equation}
displaying quantum oscillations not existing in the classical case. Once again, the amplitude of the quantum oscillations in space increases with $H$, as shown in Figs. \ref{fig8} with the stationary electron fluid density. The bunching present in the non-quasineutral case is eliminated. Similar results apply to the electron fluid velocity.

In conclusion, we have analyzed the low-frequency collisional quantum Buneman instability, both in the linear and nonlinear regimes through a one-dimensional model. Note that because this electrostatic instability is longitudinal, such a low dimensional analysis is relevant. The nonlinear evolution of the instability can be studied numerically and analytically, and results in pure quantum density oscillations.

{\bf Acknowledgments}
\vskip.3cm
This work was supported by CNPq (Conselho Nacional de Desenvolvimento Cient\'{\i}fico e Tecnol\'ogico) and by projects ENE2009-09276 of the
Spanish Ministerio de Educaci\'{o}n y Ciencia and PEII11-0056-1890 of
the Consejer\'{i}a de Educaci\'{o}n y Ciencia de la Junta de
Comunidades de Castilla-La Mancha.


\begin{thebibliography}{99}
\bibitem{Buneman1959} BUNEMAN O., {\it Phys. Rev.}, {\bf 115} (1959) 503.
\bibitem{Farley1963} FARLEY T., {\it J. Geophys. Res.}, {\bf 68} (1963) 6083.
\bibitem{Gogoberidze} GOGOBERIDZE G. {\it et al.},
{\it Astrophys. J. Lett.} {\bf 706} (2009) L12.
\bibitem{Iisuka1979} IIZUKA S., SAEKI K., SATO N. and HATTA Y. {\it Phys. Rev. Lett.}, {\bf 43} (1979) 1404.
\bibitem{Tabak1994} TABAK M., {\it Phys. Plasmas}, {\bf 1} (1994) 1626.
\bibitem{Lovelace} LOVELACE R. V. and SUDAN R. N., {\it Phys. Rev. Lett.}, {\bf 27} (1971) 1256.
\bibitem{Bret2008} BRET A. and DIECKMANN M., {\it Phys. Plasmas}, {\bf 15} (2008) 012104.
\bibitem{rmp} SHUKLA P. K. and ELIASSON B., {\it Rev. Mod. Phys.}, {\bf 83} (2011) 885.
\bibitem{usp} SHUKLA P. K. and ELIASSON B., {\it Physics-Uspekhi}, {\bf 53} (2010) 51.
\bibitem{glenzer} GLENZER S. H. and REDMER R., {\it Rev. Mod. Phys.}, {\bf 81}
(2009) 1625.
\bibitem{thiele} THIELE R. {\it et al.},
Phys. Rev. E, {\bf 82} (2010) 056404.
\bibitem{gregori} GREGORI G. and GERICKE D. O., {\it Phys. Plasmas}, {\bf 16} (2009) 056306.
\bibitem{bleker} A. SCHMIDT-BLEKER, W. GASSEN and H.-J. KULL, {\it Europhys. Lett.}, {\bf 95} (2011) 5503.
\bibitem{Ma} Y. T. MA, S. H. MAO and J. K. XUE, {\it Phys. Plasmas}, {\bf 18} (2011) 102108.
\bibitem{tzenov} TZENOV S. I. and MARINOV K. B., {\it Phys. Plasmas}, {\bf 18} (2011) 102312.
\bibitem{Stenflo} STENFLO L., SHUKLA P. K. and MARKLUND M., {\it Europhys. Lett.}, {\bf 74} (2006) 844.
\bibitem{marklund} MARKLUND M., BRODIN. G., STENFLO L. and LIU C. S., {\it Europhys. Lett.}, {\bf 84} (2008) 17006.
\bibitem{brodin} BRODIN G., MARKLUND M., ZAMANIAN J. and STEFAN M., {\it Plasma Phys. Control. Fusion}, {\bf 53} (2011) 074013.
\bibitem{tito} MENDON\c{C}A J. T., {\it Phys. Plasmas}, {\bf 18} (2011) 062101.
\bibitem{ghosh} GHOSH S. and CHAKRABARTI N., {\it Phys. Rev. E}, {\bf 84} (2011) 046601.
\bibitem{book} HAAS F., {\it Quantum Plasmas, an Hydrodynamic Approach} (Springer, New York) 2011.
\bibitem{shokri} SHOKRI B. and NIKNAM A. R., {\it Phys. Plasmas}, {\bf 12} (2005) 062110.
\bibitem{ms} HAAS F., MANFREDI G. and FEIX M., {\it Phys. Rev. E}, {\bf 62} (2000) 2703.
\bibitem{bychkov} BYCHKOV M., MODESTOV M. and MARKLUND M., {\it Phys. Plasmas}, {\bf 15} (2008) 13.
\bibitem{epl} HAAS F., {\it Europhys. Lett.}, {\bf 77} (2007) 45004.
\bibitem{pre} HAAS F., ELIASSON B., SHUKLA P. K. and MANFREDI G., {\it Phys. Rev. E}, {\bf 78} (2008) 056407.
\bibitem{Pinney} PINNEY E., {\it Proc. Am. Math. Soc.}, {\bf 1} (1950) 681.
\end{thebibliography}
\end{document}